\documentstyle [12pt,epsfig]{article} 
\makeatletter

\newcommand{\OR}{{\rm O}}
\newcommand{\ORP}{{\rm O}'}  
\newcommand{\AR}{{\rm A}}
\newcommand{\BR}{{\rm B}}  

\@addtoreset{equation}{section}
\makeatother

\newcommand{\ba}{\begin{eqnarray}}
\newcommand{\ea}{\end{eqnarray}}

\begin{document}
\newcommand{\BS}{\bigskip}
\newcommand{\SECTION}[1]{\BS{\large\section{\bf #1}}}
\newcommand{\SUBSECTION}[1]{\BS{\large\subsection{\bf #1}}}
\newcommand{\SUBSUBSECTION}[1]{\BS{\large\subsubsection{\bf #1}}}

\begin{titlepage}
\begin{center}
\vspace*{2cm}
{\large \bf Time and motion in physics: the Reciprocity Principle, relativistic invariance of
  the lengths of rulers
 and time dilatation}  
\vspace*{1.5cm}
\end{center}
\begin{center}
{\bf J.H.Field }
\end{center}
\begin{center}
{ 
D\'{e}partement de Physique Nucl\'{e}aire et Corpusculaire
 Universit\'{e} de Gen\`{e}ve . 24, quai Ernest-Ansermet
 CH-1211 Gen\`{e}ve 4.
}
\newline
\newline
   E-mail: john.field@cern.ch
\end{center}
\vspace*{2cm}
\begin{abstract}

Ponderable objects moving in free space according to Newton's First Law constitute 
 both rulers and clocks when one such object is viewed from the rest frame of
 another. Together with the Reciprocity Principle this is used to demonstrate, in both 
 Galilean and special relativity, the invariance of the measured length of a ruler
 in motion. The different times: `proper', `improper' and `apparent' appearing in different
 formulations of the relativistic time dilatation relation are discussed and exemplified
 by experimental applications. A non-intuitive `length expansion' effect predicted by
 the Reciprocity Principle as a necessary consequence of time dilatation is pointed out.


\vspace*{1cm}
\end{abstract}
\end{titlepage}
 
\SECTION{\bf{Introduction}}
 The standard text-book presentation of special relativity follows closely that of Einstein's
 seminal paper of 1905~\cite{Ein1} in basing the theory on the Special Relativity Principle,
 classical electromagnetism and the postulate of constant light speed. However an alternative
  and conceptually
   simpler approach to the physics of space and time, in the absence of gravitational
   fields, is possible in which it is not necessary to consider light signals, classical
  electromagnetism, or indeed, any dynamical theory whatsoever. The Lorentz transformation (LT)
   was first derived in this way by Ignatowsky~\cite{Ignatowsky} in 1910. Purely mathematical
   considerations imply, in such a derivation of the LT, the existence of a maximum relative
   velocity, $V$, of two inertial frames. Use of relativistic kinematics then shows
   that $V$ is equal to the speed of light, $c$, when light in identified as a manifestation
   of the propgation in space-time of massless particles --photons~\cite{JHFLT1}. In this way
   Einstein's mysterious second postulate is derived from first principles. The fundamental
  axiom underlying such an approach is the Reciprocity Principle (RP)~\cite{BG,JHFLT1},
  discussed in Section 3 below, relating the the relative velocities of two inertial frames.
    Derivations of the LT and the parallel velocity addition formula based on the RP
    and other simple axioms are given in Ref.~\cite{JHFLT1}. 
   \par In the present paper the space-time properties of ponderable\footnote{
     That is bodies, with a non-vanishing Newtonian mass, which may be associated 
   with an inertial frame in which the body is at rest. No such frame may be
    associated with a massless object.} physical bodies in free
    space, as described by Newton's First Law of mechanics, are used together with the RP,
  to demonstrate the invariance of the measured length of a ruler in uniform motion.
   The proof given is valid in both Galilean and special relativity, since Newton's First Law
    and the RP hold in both theories. 
   \par The analysis presented is based on a careful definition of physical time concepts.
       In particular, the `frame time' or `proper times' that appear in in the RP,
     are distinguished from the `improper time' or `apparent time' (of a moving clock) that
     appear in the Time Dilatation (TD) relation of special relativity. 
    \par The paper is organised as follows: The following section contains 
   an elementary discussion of the concepts of `space', `time' and `motion' in physics,
   in connection with Newton's First Law. In Section 3, the RP is introduced
    and discussed in relation to Newton's First Law. It is pointed out that,
    because of the RP, `rulers are clocks' and `clocks are rulers' when the motion
   of ponderable bodies in free space is considered. In Section 4 the RP is
   used to demonstrate the invariance of the measured length of a uniformly
   moving ruler. In Section 5 the operational meanings of the time symbols
    appearing in the TD formula of special relativity are discussed. This may be done in
    a `clock oriented' manner in terms of 'proper' and 'improper' times of the observed
    clock, or in an `observer oriented' manner in terms of the proper time of the
     observer's local clock and the `apparent time', as seen by the observer,
     of the moving clock. Two specific
     experiments are described to exemplify the operational meanings of the time symbols
    in the TD formula. A non-intuitive `length expansion' effect is found to relate 
    similarly defined spatial intervals corresponding to the observation of an event
    either in the rest frame of the clock, or in a frame in which it is in uniform 
    motion. 
    \par The results of the present paper show that the `length contraction' effect
      and the correlated `relativity of simultaneity' effect of conventional 
      special relativity do not exist. A detailed discussion of the reason for the spurious
     nature of these effects of conventional
       special relativity theory may be found in
       Refs.~\cite{JHFLLT,JHFST1,JHFUMC,JHFCRCS,JHFAS,JHFCOORD} .
      \par However, a genuine `relativistic length contraction' effect does occur
       when distances between spatial coincidences of moving objects 
       are observed from different inertial frames~\cite{JHF2TEXP}. Also a genuine
     `relativity of simultaneity' effect occurs when clocks at rest in two different
   inertial frames are viewed from a third one~\cite{JHFTETE,JHFMUDEC}. An 
    alternative derivation, directly from the RP, of the invariance
    of the measured spatial separation of two objects at rest in the same inertial
      frame as well as the absence of the conventional `relativity of simultaneity'
      effect is given in Ref.~\cite{JHFAS}.

\SECTION{\bf{Physical time and Newton's First Law of Mechanics}}

 In physics the concepts of `time' and `motion' are inseparable. In a world in which
 motion did not exist the physical concept of time would be meaningless. Similarly
 the physical concepts of `space' and `motion' are inseperable. Without the concept
 of space, no operational definition of motion is possible. The concept of historical
 time --the time of the everyday world of human existence-- requires the introduction
  of the further, and equivalent, concepts of `uniform motion' and `cyclic
   motion with constant period'. For example, the unit of time the `year' is identified
  with the (assumed constant) period of rotation of the Earth about the Sun. 
  \par The idea of uniform motion entered into physics in a quantitative way with the
   formulation of Newton's First Law~\cite{CW}
  \par {\tt Every body continues in its state of rest, or uniform motion in a right \newline
   line unless it is compelled to change that state by forces impressed upon \newline it. }
   \par This law gives an operational meaning to the physical concept of `uniform motion'.
      It is defined by observations of the position of any ponderable object in
    `free space' i.e. in the absence of any mechanical interaction of the object with other
     objects. There is a one-to-one correspondence between such a ponderable object and an
    `inertial frame' of relativity theory. As will be discussed in the following section,
     one such ponderable object, O, constitutes both a ruler and a clock for an observer
      in the rest frame of another such object, O', and {\it vice versa}. 
   \par When time is measured by using a cyclic physical phenomenon, e.g. an analogue clock,
     time measurement reduces to recording the result of a spatial (or angular) measurement.
     There is a one-to-one correspondence between the spatial coincidence of a 
    stationary `mark' on the face of the clock and a moving `pointer', constituted by the
    hand of the clock, and the time measurement~\cite{JHFST1}. A `time interval' is measured
    by the angular separation of two such `pointer-mark coincidences'. The implicit
    assumption is that the motion of the pointer is `uniform'. There is an evident
   logical circularity here since `equal' time intervals measured by such an analogue clock
    assume that the angular velocity of the hand is constant, whereas constant angular velocity
   is established by observation of equal angular increments for equal time time intervals
   (i.e. also equal angular increments) recorded by a second clock of supposedly known uniform rate.
    In practice, this
    conundrum is resolved by an appeal to physics. For example, an undamped pendulum in a uniform
   gravitational field is predicted, by the laws of mechanics, to have a constant period
   of oscillation. Quantum mechanics predicts the same transition frequency and mean lifetimes
   for two identical atoms in the same excited state, in the same physical environment, etc.
   \par Measurements of `time' are then ultimately observations of spatial phenomena, e.g.
   the time measurement corresponding to observation of the number displayed by a digital clock 
   is a spatial perception. This will also be the case
     for time measurements related to observation of two ponderable objects O and O' in motion 
   in free space that will now be discussed.

\SECTION{\bf{The Reciprocity Principle: rulers are clocks, and clocks are rulers}}
  Consider two non-interacting ponderable objects O and O', with arbitary motions in 
  free space. They are placed at the origins of inertial coordinate systems S and S' 
 with axes orientated so that the $x$ and $x'$ axes are parallel to the relative 
   velocity vector of O and O'. Without any loss of generality for the following
   discussion, it may be assumed that  O and O' lie on the common  $x$-$x'$ axis.
  \par The Reciprocity Principle (RP)~\cite{BG,JHFLT1,JHFAS} is defined by the equation:
    \begin{equation}
        v = v_{\ORP\OR} = \left.\frac{\partial x_{\ORP\OR}}{\partial t}\right|_{x'_{\ORP}x_{\OR}}
           =  -\left.\frac{\partial x'_{\OR\ORP}}{\partial t'}\right|_{x_{\OR}x'_{\ORP}}
         = -v_{\OR\ORP}
    \end{equation}
    where $ x_{\ORP\OR} \equiv x_{\ORP}-x_{\OR}$ and $x'_{\OR\ORP} \equiv x'_{\OR}-x'_{\ORP}$, or in words:
    `If the velocity of O' relative to O is $\vec{v}$, the velocity of O relative to O' is - $\vec{v}$'.
      In many discussions of special relativity, the RP is taken as `obvious' and is often not even
      declared as a separate axiom. This is the case, for example, in Einstein's 1905
      special relativity paper~\cite{Ein1}. However, as first demonstrated by Ignatowsky in
      1910~\cite{Ignatowsky}, it is sufficient, together with some other weaker axioms such as
      the homogeneity of space or single-valuedness of the transfomation equations, to
      derive~\cite{JHFLT1} the space-time Lorentz transformation and hence the whole
        of special relativity theory. 
      \par Eqn(3.1) looks very similar to the equation defining the relative velocity of two
      objects A and B as observed in a {\it single} inertial reference frame (say S):
     \begin{equation}
        v_{\AR\BR} \equiv v_{\AR}- v_{\BR} = \frac{d ( x_{\AR}- x_{\BR})}{d  t} \equiv       
          \frac{d x_{\AR\BR}}{d  t}  =  - \frac{d x_{\BR\AR}}{d  t} = - v_{\BR\AR}
    \end{equation}
      The crucial difference is the appearence in the RP, (3.1), of 
     {\it two different times} $t$ and $t'$. The time $t$ is the `frame time' of S. i.e. the time
     registered by a synchronised clock at rest, at any position in S, according to
     an observer also a rest in S. The frame time 
     $t'$ is similarly defined by an array of synchronised clocks at rest in S'.
      Eqn(3.1) (and its integral) gives a relation between the times $t$ and $t'$ 
      Both $t$ and $t'$ correspond to `proper times' of clocks at rest, whereas, as
      explained in Section 4
     below, the Lorentz transformation relates instead a proper time to an `improper time'
     --the observed time of a clock in uniform motion.
      \par Suppose now that O and O' are equipped with local clocks that are observed
    to run at exactly the same rate when they are both at rest in the same inertial 
    frame. The direction  of the relative velocity vector $\vec{v}$ of O' relative
    to O is such that they are approaching each other at the frame times $t$ and $t'$.
    The spatial separations of O and O' in S and S' are $\ell(t)$ and $\ell'(t')$
    respectively, at times  $t$ and $t'$. Using the RP, a spatial coincidence of
     O and O' will be observed at the time
     \begin{equation}
      t_{\OR\ORP} = t +\frac{\ell(t)}{v}
  \end{equation}
    in S, and
     \begin{equation}
      t'_{\OR\ORP} = t' +\frac{\ell'(t')}{v}
  \end{equation}      
   in S'. The OO' coincidence event will be {\it mutually simultaneous} in the frames 
   S and S'.
   \par Note that the OO' spatial coincidence that is mutually simultaneous in S and S' constitutes
    a pair of {\it reciprocal} pointer mark coincidences. In S the mark is at the position of O
    and the moving pointer at the position of O', whereas in S' the position of O' constitutes
    the mark and the position of O the pointer. A corollary is that all such pairs of
     reciprocal pointer mark coincidences are mutually simultaneous. This is the basis of
    the `system external synchronistation'~\cite{MS} as introduced in Einstein's first
    special relativity paper~\cite{Ein1} to synchronise clocks at rest in different inertial frames
    when they are in spatial coincidence.
    \par The observation of the OO' coincidence event in both frames can be used to give a
   condition that {\it any} other pair of events, one observed in S, the other observed in S'
   are mutually simultaneous. If the time of an event in S is $\tilde{t}$ and another event in S' is
    $\tilde{t}'$ they will be `mutually simultaneous' providing that:
     \begin{equation}
  \tilde{t}'-\tilde{t} =  t'_{\OR\ORP}- t_{\OR\ORP}
   \end{equation} 
   Combining (3.3)-(3.5) gives:
      \begin{equation}
    \tilde{t}'-\tilde{t} =  t'_{\OR\ORP}- t_{\OR\ORP} = t'- t +\frac{\ell'(t')- \ell(t)}{v}
   \end{equation} 
       If now events occuring at times $t$ in S and  $t'$ in S' are mutually simultaneous,
    it follows from (3.5) and (3.6) that $ \ell(t) = \ell'(t')$, so that 
      {\it events which occur when O and O' have the same spatial separation in S and S' are 
      mutually simultaneous}. A special case occurs if the clock arrays in  S and S' 
      are {\it mutually synchronised} so that  $ \ell(t) = \ell'(t' = t)$. There is then 
     a direct correlation between either $t$ or $t'$ and the spatial separation of O and O':
     {\it When mutually synchronised clocks in the frames S and S' have the same reading, O and O'
      have the equal spatial separations in S and S'}, and conversely, {\it When O and O' have
      equal spatial separations in the frames S and S', mutually synchronised clocks in
      S and S' have the same reading}.
      \par The dependence of  $\ell$ on $t$ in Eqn(3.3) and $\ell'$ on $t'$ in Eqn(3.4)
      means that each of the objects may be considered to be an `inertial clock' 
       by an observer in the rest frame of the other one. That is, $t$ is  measured  by the
        spatial separation of O' from O in S and $t'$ is measured by the spatial separation
      of O from O' in S'. Conversely, after mutual synchronisation of the
       clock arrays in S and S' at the instant when O and O' are
       in spatial coincidence, $t$ measures the spatial separation of O' and O in S
       (and so is effectively a ruler in this frame) while $t'$ measures the spatial
      separation of O' and O in S', constituting a ruler in this frame.
       Matching of these measurements of the separation of O and O' with
    the lengths of physical rulers at rest in S and S' is now used to demonstrate the
   invariance of the measured length of the length of a
   ruler in uniform motion --that is, the absence of any relativistic length contraction 
     effect-- in this case.

\SECTION{\bf {Invariance of the measured length of a ruler in uniform motion}}

\begin{figure}[htbp]
\begin{center}\hspace*{-0.5cm}\mbox{
\epsfysize10.0cm\epsffile{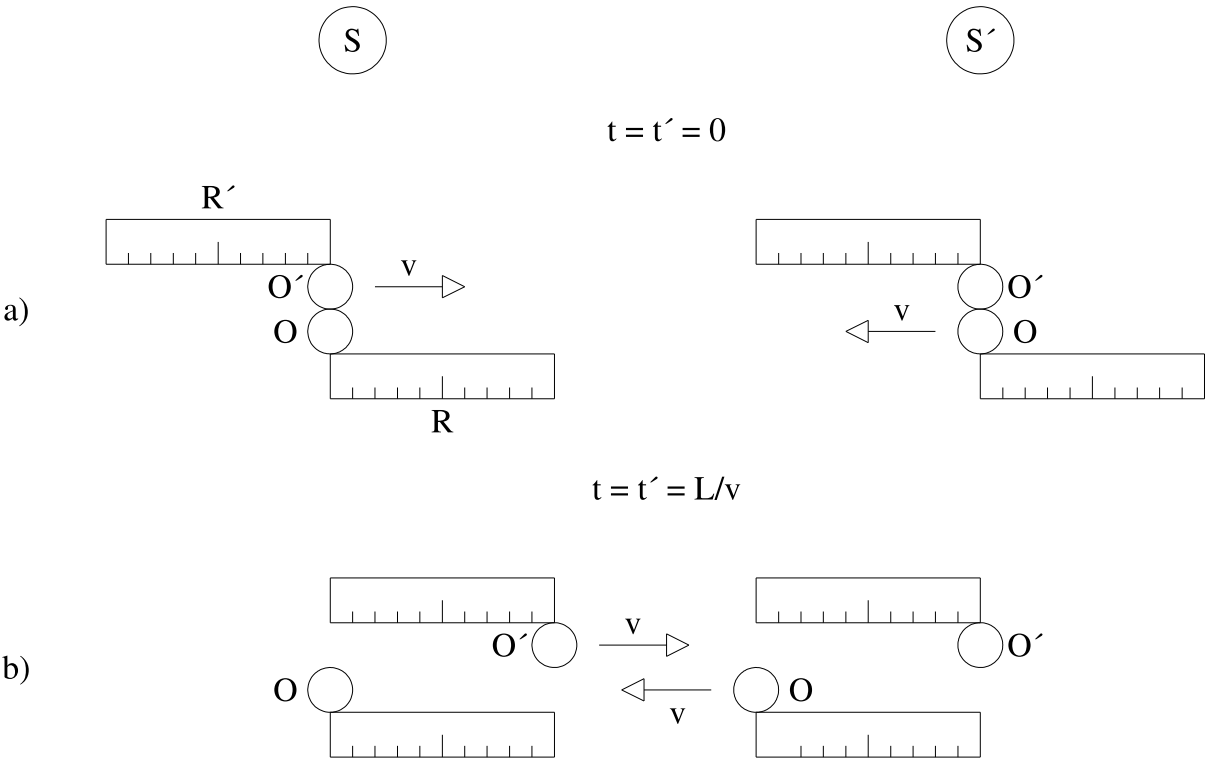}}
\caption{{\em Rulers attached to objects O and O' are viewed from the frame S (left) and
  S' (right). The equality of the separations of O and O' in S and S' at time
     $t = t' = L/v$, predicted by the RP, is used to establish the invariance of the measured
       length of the moving ruler R' in S, or of the moving ruler R in S' (see text).}}
\label{fig-fig1}
\end{center}
\end{figure}

 Suppose that O and O' are equipped with rulers R and R', parallel to the $x$-$x'$ axis as shown
 in Fig.1. O coincides with the mark $MR(0)$ of the ruler R and O' with the mark $MR'(10)$
    of the ruler R'. A $t = t' = 0$ (Fig.1a) O and O' are in spatial coincidence. The clock
    arrays in S and S' are mutually synchronised at this time.
     The length of each ruler in its rest frame is $L$. The object O' now moves along the 
    ruler R, being in spatial coincidence with different marks of the ruler at different
    times. The object O moves in a similar manner along the ruler R'. At any given
    time $t$ the separation of O and O' in S is given by the corresponding `Pointer
    Mark Coincidence' (PMC):
    \begin{equation}
     PMC(\ORP,t)\equiv \ORP(t)@MR(J)  
    \end{equation}
     where the symbol before the ampersand denotes the moving `pointer', and the symbol
     after it the stationary `mark' with which it is spatial coincidence\footnote{This notation 
     was introduced in Ref.~\cite{JHFST1}. Note the similarity with an e-mail address}.  
     Since
  \begin{equation}
     PMC(\OR,t)\equiv \OR(t)@MR(0)~~~~~ {\rm for~all}~t   
    \end{equation}
   and $x[MR(0)] = 0$ it follows that the separation of O and O' in the frame S at time $t$
   is given by:
    \begin{equation}
     d_{\ORP\OR}(t) = x[MR(J)]-x[MR(0)] = x[MR(J)] 
    \end{equation}
     where
 \[  x[MR(J)] = \frac{J L}{J_{max}} \]
  and where, in Fig.1, $J_{max} = 10$, is the ordinal number of the mark at the end of
  the ruler. Thus the $x$-coordinate origin is at $MR(0)$.
   Defining in a similar manner a PMC in the frame S':
     \begin{equation}
     PMC(\OR,t')\equiv \OR(t')@MR'(K)  
    \end{equation}
   and since
  \begin{equation}
     PMC(\ORP,t')\equiv \ORP(t')@MR'(10)~~~~~ {\rm for~all}~t'   
    \end{equation}
  the separation of O and O' in S' at the time $t'$ is 
  \begin{equation}
     d'_{\OR\ORP}(t') = x'[MR'(10)]-x'[MR'(K)] 
    \end{equation}
     where
  \[  x'[MR'(K)] = \frac{K L}{K_{max}} \]
    and where, in Fig.1, $K_{max} = 10$. 
   The spatial configurations in S and S' at the times
    $t = t' = L/v$ are shown in Fig.1b. The corresponding PMC
    are:
  \begin{eqnarray}   
    PMC(\ORP,L/v) & \equiv & \ORP(L/v)@MR(10) \\
    PMC(\OR,L/v) &\equiv & \OR(L/v)@MR'(0)    
   \end{eqnarray}
    It follows from (4.3) and (4.6) that
    \begin{equation}
     d_{\ORP\OR}(L/v) = x[MR(10)]- x[MR(0)] = L =  x'[MR'(10)]-x'[MR'(0)] =  d'_{\OR\ORP}(L/v) 
    \end{equation} 
     Since O' coincides with $MR'(10)$ at all times it follows that, at
     $t = L/v$
    \begin{equation} 
     x[MR'(10)] = x[\ORP] = x[MR(10)]
     \end{equation}
    Also, since O is in spatial coincidence with $MR'(0)$ at $t = t' = L/v$ it follows
    that at  $t = L/v$, 
      \begin{equation} 
     x[MR'(0)] = x[\OR] = x[MR(0)] = 0
     \end{equation}
    Eqns(4.9)-(4.11) then give at  $t = L/v$:
  \begin{equation} 
     x[MR'(10)]- x[MR'(0)] = x[MR(10)]- x[MR(0)] = L       
  \end{equation}
    That is, the measured length of the moving ruler R' in the frame S, at  $t = L/v$, is the same
    as the length of the same ruler at rest --there is no `length contraction' effect.
   A similar calculation for the length of the ruler R as measured in the 
  frame S' gives, at  $t' = L/v$:
 \begin{equation} 
     x'[MR(10)]- x'[MR(0)] = x'[MR'(10)]- x'[MR'(0)] = L       
  \end{equation}
   The length of the moving ruler R as measured in S', at  $t' = L/v$, is the same as the length of 
    the same ruler at rest. The above calculations 
    have used the equality of the spatial separations of O and O' in S and S'
    at equal times of mutually synchronised clocks in these frames, that follows from the RP,
    to establish, via corresponding PMCs, the equality of the measured lengths
    of a ruler at rest, or in motion. Note that nowhere in any of the calculations
     was the Lorentz transformation invoked. In fact the calculations are the same in
     Galilean and special relativity, since the RP is equally valid for both.

\SECTION{\bf {The time dilatation effect; proper, improper and apparent time intervals}}

\begin{figure}[htbp]
\begin{center}\hspace*{-0.5cm}\mbox{
\epsfysize9.0cm\epsffile{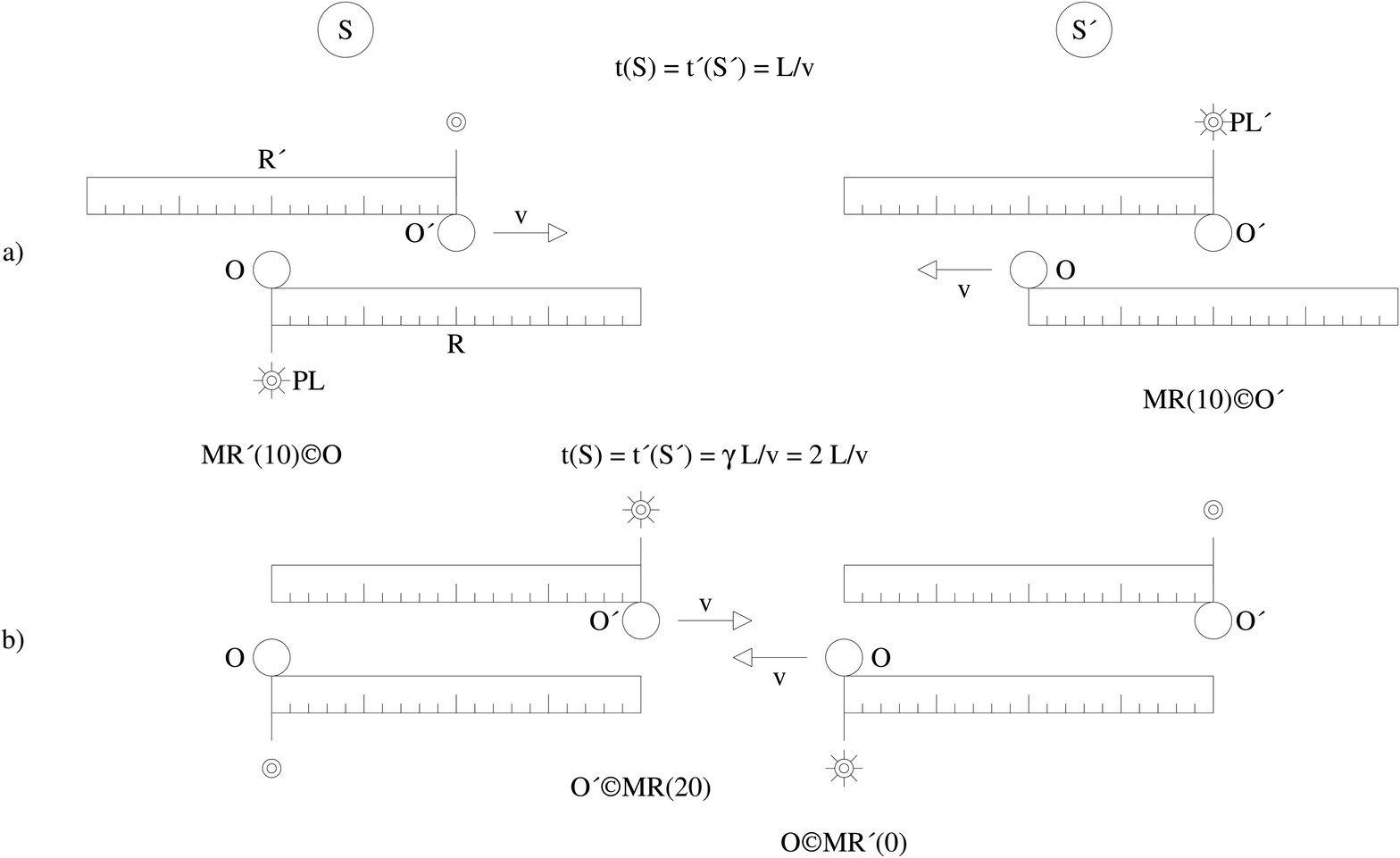}}
\caption{{\em An experiment to illustrate the TD effcet viewed from S (left) and S' (right).
 a) The pulsed lamp PL at rest in S flashes at time $t(S) = L/v$ and PL' at rest in S'
   flashes at time $t'(S') = L/v$. b) The light signal from PL is observed at time 
   $t(S') = \gamma L/v$ in the frame S', that from PL' at time  $t'(S) = \gamma L/v$
     in the frame S. The PMCs corresponding to the positions of observation of the
 signals in the different frames are indicated. See text for discussion.}}
\label{fig-fig2}
\end{center}
\end{figure}
\begin{figure}[htbp]
\begin{center}\hspace*{-0.5cm}\mbox{
\epsfysize16.0cm\epsffile{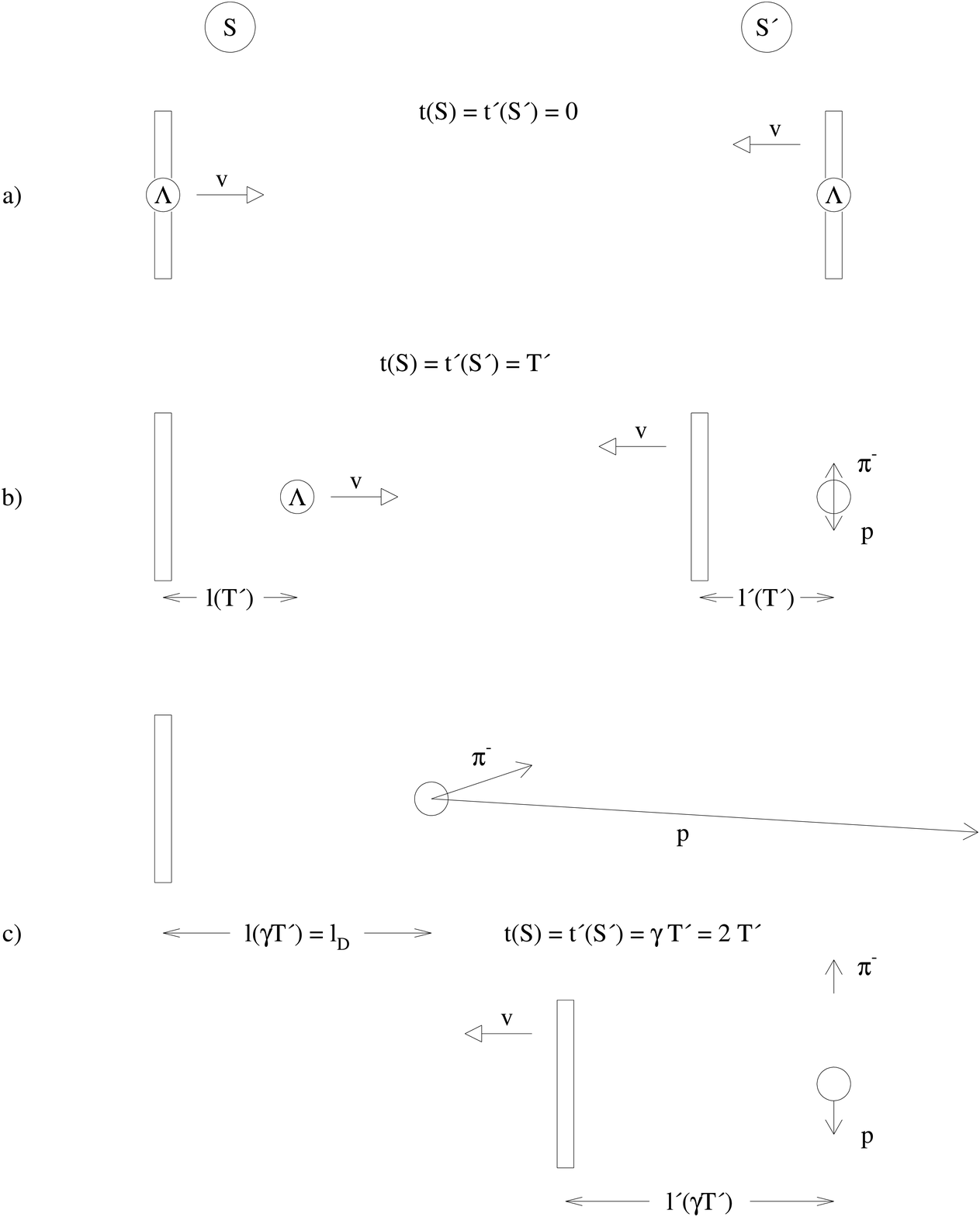}}
\caption{{\em Spatial configurations in the frame S (left) and the frame S' (right)
   are viewed at different times. a) $t(S) = t'(S') = 0$; the $\Lambda$ is created 
    and moves to the right in the plane of the figure with speed $v = \sqrt{3}c/2$.
   b) $t(S) = t'(S') = T'$; the  $\Lambda$ is observed to decay in the frame S'.
    The decay products move in the plane of the figure perpendicular to the direction
     of motion of the $\Lambda$. c) $t(S) = t'(S') = \gamma T'$; the  $\Lambda$ is observed
      to decay in the frame S.
     See text for discussion. The momentum vectors of the $p$ and $\pi^-$ are drawn to
     scale in the different reference frames. The spatial position of each particle is at the tail
      of the corresponding momentum vector.}}
\label{fig-fig3}
\end{center}
\end{figure}
    All the times considered above were `frame times' i.e. $t$ and $t'$ are the times
   recorded by a synchronised clock at rest at any position in S and S' as viewed
   by an observer at rest in these respective frames. In order to discuss the time dilatation
   effect it will be found convenient to use the notation $t(S)$, $t'(S')$ for the frame times
   where the arguments S, S' specify the reference frame of the observer of the clock.
  Such times are {\it proper times} of such a clock. The Lorentz transformation
  relates the space-time coordinates ($x'$,$t'(S')$) of an event specified in the frame S' 
  to those of the {\it same event}, ($x$,$t'(S)$) as observed in S, or {\it vice versa}.
  The times  $t(S')$[ $t'(S)$] which are those of clocks at rest in S[S'], as viewed
  from S'[S] are called {\it improper times}. The space-time LT gives the 
  following invariant interval relation between corresponding space and time
  intervals in the frames S and S':
  \begin{equation} 
  c^2 (\Delta \tau')^2 = c^2 (\Delta t'(S))^2-(\Delta x)^2 = c^2 (\Delta t'(S'))^2 -(\Delta x')^2
  \end{equation}
  where $\Delta x \equiv x_2-x_1$ etc, while the inverse LT gives:
  \begin{equation} 
  c^2 (\Delta \tau)^2 = c^2 (\Delta t(S'))^2-(\Delta x')^2 = c^2 (\Delta t(S))^2 -(\Delta x)^2
  \end{equation} 
   In order to use the general interval relation (5.1)
   to derive the time dilatation effect it is necessary to identify the time 
   interval $\Delta t'(S')$ with the proper time interval of a clock at rest in S'
   ($\Delta x' = 0$), and with equation of motion in S:  $\Delta x = v \Delta t'(S)$.
   Using the latter equation to eliminate $\Delta x$ from (5.1) and setting $\Delta x' = 0$
   yields the time dilatation (TD) relation:
 \begin{equation} 
 \Delta t'(S) = \gamma \Delta t'(S')
 \end{equation}
  where $\gamma \equiv 1/\sqrt{1-(v/c)^2}$, relating the improper to the proper time of a
   clock at rest in S'.
   In a similar manner the interval relation (5.2) gives the TD relation for a clock
   at rest in S and observed from S':
 \begin{equation} 
 \Delta t(S') = \gamma \Delta t(S)
 \end{equation}
  It is important to note the existence of four different time symbols, with different
  operational meanings in Eqns(5.3) and (5.4). The proper times $t(S)$ and $t'(S')$
  (corresponding to the `frame times' $t$ and $t'$ of the previous sections) and the
   improper times  $t(S')$ and $t(S')$. The notation for these times just introduced
    may be called `clock oriented' since only the readings of a single clock
    (observed either at rest, or in motion) appear in the TD relations. In any
   actual experiment where the TD effect in measured, two clocks are necessary,
   the observed moving clock, and another one at rest to measure the corresponding time interval
    in the observer's proper frame. If a clock at rest in S' is observed from S as 
    in Eqn(5.3), the time interval $\Delta t'(S)$ is actually that, $\Delta \tau$,
    recorded by a similar clock, at rest in S while $\Delta t'(S')$ is the corresponding
    time interval recorded by the (slowed-down) moving clock. Since the observed rate
   of the moving clock depends on its motion,  $\Delta t'(S')$ is not a proper time interval
    for the observer in S. From the view-point of the latter this is an `apparent'
     (velocity-dependent) time interval that may be denoted simply as  $\Delta t'$,
     to distinguish it from the observer's proper time interval  $\Delta \tau$. This
    gives an alternative `observer oriented' time notation for the TD relations
    (5.3) and (5.4) above:
     \begin{eqnarray}
 \Delta \tau & = & \gamma \Delta t' \\
\Delta \tau' & = & \gamma \Delta t
   \end{eqnarray}
  This alternative notation has beeen employed in several previous papers by
  the present author~\cite{JHFST1,JHFCRCS,JHF2TEXP,JHFTETE,JHFMUDEC,JHFST2}. 

  \par In order to apply the TD relations (5.3) and (5.4), or (5.5) and (5.6), to
   any actual or imagined experiment an operational definition must be given
    to the improper time intervals of Eqns(5.3) and (5.4) or the apparent
    time intervals of (5.5) and (5.6). Two examples of such definitions will
    be given, the first in a thought experiment to illustrate the physical
    meaning of the TD effect, the second in an actual experiment
   typical of many performed in particle physics, where the TD effect is
   used to measure the proper decay time of an unstable particle. However
    as will be seen, the thought experiment and actually realisable (and many times
    realised) one are similar in all essential features.

  \par What notation is most convenient depends on the experiment considered.
   In the observation of the TD effect in the last CERN muon g-2 experiment~\cite{MUGM2}
     where the time interval $\Delta \tau$ was directly measured by clocks in the 
     laboratory frame, and $\Delta t'$ was the known muon rest-frame lifetime,
     it was natural to use Eqn(5.5). For the second of the two experiments considered below where
  $\Delta \tau$ is not directly measured but inferred from spatial measurements in the frame
   S, the relation (5.3) relating connecting a proper time in the frame S' to an improper time
   in the frame S, is used.  

    \par In the thought experiment it is imagined that the objects O, O' are each equipped
    with local pulsed lamps PL, PL'. The objects O, O' are in spatial coincidence
    at times $t(S) = t'(S') = 0$ and are attached to  rulers of length $2L$ in similar
    spatial configurations to that shown in Fig.1a. The objects move apart with 
    relative velocity $v = \sqrt{3}c/2$. As shown in Fig.2a, at the times
     $t(S) =t'(S') = L/v$, PL and PL' both flash, producing an isotropic pulse
    of photons. The observation times in S of the photon signal produced by PL',
    and in S' of the photon signal produced by PL, are given by Eqns(5.3) and (5.4)
    respectively. Since $\gamma = 2$, these observations occur at the times 
    $t(S) =t'(S') = \gamma L/v = 2 L/v$. The corresponding spatial configurations
    of O and O' at these times shown in Fig.2b. It can be seen that the observation
    times of the light flashes in S and S' correspond to different PMCs of the
     objects O and O' and to different spatial separations of the objects:
    \begin{eqnarray}
      {\rm In~S~~~PL:~~~} PMC(MR'(10),L/v) & \equiv & MR'(10)@0 = MR'(10)@MR(0)  \\
   {\rm PL':~~~~~~~~~} PMC(\ORP,\gamma L/v) & \equiv & \ORP@MR(20) = MR'(20)@MR(20) \\
      {\rm In~S'~~~PL':~~~} PMC( MR(10),L/v) & \equiv & MR(10)@O' = MR(10)@MR'(20) \\
   {\rm PL:~~~~~~~~~~} PMC(\OR,\gamma L/v) & \equiv & \OR@MR'(0) = MR(0)@MR'(0)
\end{eqnarray}
  \begin{equation}
  \frac{\ell(\gamma L/v)}{\ell(L/v)} = \frac{\ell'(\gamma L/v)}{\ell'(L/v)} 
     = \frac{v\gamma L/v}{vL/v)} = \gamma
  \end{equation}
 The relations in (5.11) follow directly from the RP, while the PMC in (5.7)-(5.10)
  are obtained from the geometry of Fig.2 and the invariance of the lengths of the
   moving rulers derived in Section 3 above. 
    \par The different PMC corresponding to observations of the light flashes emitted by
   PL and PL' in different
  frames in (5.7)-(5.10) is deeply perplexing for common-sense concepts of space and 
  time. For example the photon bunches emitted by PL' correspond to $MR(10)@MR'(20)$ in S'
   and to $MR'(20)@MR(20)$ in S. In some discussions of time dilatation this apparent
   paradox is avoided by invoking a hypothetical contraction of a moving ruler by
   a factor $1/\gamma$~\cite{NDMIAT}. This has the effect of shortening the moving
   ruler R by a factor $1/2$ in the right hand figure in Fig.2a, so that the PMC
   corresponding to the flashing of PL' becomes
    $MR(20)@MR'(20)$, the same as in S with inversion of pointer and mark. However,
    as demonstrated above, there is no such length contraction effect, which, as pointed
    out elsewhere~\cite{JHFLLT,JHFST1,JHFUMC,JHFCRCS,JHFAS}
    is a spurious consequence of misinterpreting
    the space-time Lorentz transformation. Indeed the possibility of such a length
   contraction effect is already excluded by inspection of Fig.2a. In the
    right hand figure, the PMC correponding to the moving object O considered as 
    a pointer is $MR(0)@MR'(10)$. Since O is in motion and R' at rest no hypothetical
    length contraction effect operates here. In the left hand figure the mutually
   simultaneous PMC in S is $MR'(10)@MR(0)$ so that at $t(S) = t'(S') = L/v$ observers
    in S and S' see reciprocal PMCs, i.e. ones related by exchange of the pointer
    and mark symbols. If however the length contraction effect exists, the observer in
     S will see instead that the PMC corresponding to O is  $MR'(0)@MR(0)$ 
     at time $t(S) = L/v$. But from the RP this PMC must correspond to the times 
     $t(S) = t'(S') = 2 L/v$ (see Fig.2b) contrary to the assumption that
      $t(S) = L/v$. The length contraction hypothesis therefore contradicts
     the corollory of the RP that states that mutually simultaneous events
     in two frames have reciprocal PMCs, since it implies that the reciprocal
     PMCs  $MR'(0)@MR(0)$ and $MR(0)@MR'(0)$ are not mutually simultaneous.

      \par The second example of a TD experiment illustrates a typical application
       of the effect in particle physics (see Fig.3). A $\pi^-$ meson interacts
      with a proton in a thin plastic target T to produce a $\Lambda$ hyperon via the
      reaction\footnote{The results of an actual such
       experiment constructed to test the $\Delta S = \Delta Q$ rule in
      semileptonic neutral kaon decays are described in Ref.~\cite{JHetal}.} $\pi^-p\rightarrow\Lambda K^0$          The hyperon moves with 
      velocity $v = \sqrt{3}c/2$ perpendicular to the plane of the target
       in the laboratory frame S. After the time $t'(S') = T'$ in its rest
       frame S', it decays to a proton and a negative pion:
       $\Lambda \rightarrow p \pi^-$. These decay products are observed in
       the laboratory system. The experiment is in every way similar to that shown
       in Fig.2. The object O is replaced by the target T, the object O' by
     the undecayed  $\Lambda$ or the kinematical system constructed from its
     decay products. The photon pulse emitted by PL' is replaced
      by the decay products of the $\Lambda$. By reconstructing the
      trajectories of the decay $p$ and $\pi^-$ in a particle detector the position of the
       decay event and hence the decay length $l_D$ --the distance between the
       point of production and decay of the  $\Lambda$-- in the frame S can be measured.
       Identification of the $p$ and $\pi^-$ and measurement of their momenta
       (typically by measurement of the curvature of their trajectories in 
       a known magnetic field ) enables the momentum $P$ and the energy $E$ of
        the $\Lambda$ to be determined. Since $v = Pc^2/E$ and $\gamma = E/(m_{\Lambda} c^2)$
       where $m_{\Lambda}$ is the mass of the $\Lambda$, the proper decay time of the 
       $\Lambda$ is given by Eqn(5.3) as:
       \begin{equation}
        T' = \Delta t'(S') = \frac{\Delta t'(S)}{\gamma} = \frac{l_D}{\gamma v}
           = \left(\frac{m_{\Lambda} c^2}{E}\right)\left(\frac{E}{Pc^2}\right)l_D
               = \frac{m_{\Lambda}l_D}{P} 
         \end{equation}
     The spatial configurations of T and the $\Lambda$ at different times in
      the frames S and S' are shown in Fig.3. The spatial separations of T and the $\Lambda$
      at the observed instant of decay in S and S' obey the relation (5.11). This implies that this
     separation, in changing the frame of observation from the rest frame of the
       $\Lambda$ to the laboratory system in which it is motion, undergoes a 
      `length expansion' by the factor $\gamma$. In accordance with Eqn(5.11),
       it can be seen that this is a necessary consequence of the RP, given the existence
         of the TD effect.
      \par The mutally simultaneous events in S and S' shown in Fig.3c, correspond, as they must,
        to equal spatial separations of T and the physical object constituted by the
      decay products, $p$ and $\pi^-$,  of the $\Lambda$. However, in the frame S, these particles
     have just been created and have vanishing spatial separation, whereas in S' they are spatially
     separated by a distance corresponding to a time-of-flight $(\gamma-1)T'$. This also seems 
     highly paradoxical when interpreted by commonsense classical concepts of space and time.
     \par {\bf Acknowledgement} 
      \par  I thank the referee of the journal that rejected Ref.~\cite{JHF2TEXP}
        for publication for correspondence that was important for the clarificatiion
        of the ideas expressed in both the latest version of Ref.~\cite{JHF2TEXP} and
        the present paper.
  \par {\bf Added Note}
     \par The calculations presented in the present paper are flawed by a major conceptual
     misunderstanding which is rectified in later papers~\cite{JHFPRSTE,JHFSTLAWS} treating 
      similar subjects.
     \par At the time of writing the present paper, the author had correctly understood the spurious
      nature of the `relativity of simultaneity'and `length contraction' effects of conventional
      special relativity theory~\cite{JHFLLT,JHFUMC,JHFCRCS,JHFCOORD} but had not yet drawn the simple conclusion that the existence
       of the genuine and experimentally-confirmed time dilatation effect then necessarily implies that the
       Reciprocity Principle, as generally understood, also breaks down in special relativity. 
        This point is easily understood by considering the first member of Eqn(3.1), written in a simplified
        notation as:
        \[ v \equiv\frac{dx_{\ORP\OR}}{dt} \]
       Transforming into the frame S', the invariance of length intervals implies that
  
        \[dx_{\ORP\OR} = -dx'_{\OR\ORP} \] Since the time dilatation relation gives
       $dt = \gamma dt'$, the Reciprocity Principle of (3.1) is replaced by:
       \[ v \equiv \frac{dx_{\ORP\OR}}{dt} = -\frac{1}{\gamma}  \frac{dx'_{\OR\ORP}}{dt'} \]
         so that 
        \[ v' \equiv \frac{dx'_{\OR\ORP}}{dt'} = -\gamma v \]
        to be compared with $v' = -v$ given by (3.1).
         \par The detailed calculations presented in Section 4 are correct and logically coherent
          given the initial assumptions, but the configurations shown in the frame S' in Fig.1 do not
         correspond to observations in this frame of the coincidence events specified in the frame S
        in the same space-time experiment. If this were the case, in the S' frame configurations
       in Fig.1 $v$ should be replaced by $\gamma v$ and $t$ and $t'$ should be related by time dilatation
         relation
        $t = \gamma t'$.  In fact, what are shown in Fig.1 and considered in Section 4 are the
         configurations in S of a primary experiment and in S' of the corresponding but {\it physically
         independent} reciprocal experiment~\cite{JHFPRSTE,JHFSTP3}. 
         \par Nevertheless, the invariance of corresponding length intervals can be derived~\cite{JHFSTLAWS}
           by considering the configurations in S and S' in Fig.1b in the case that they
           are corresponding ones, at the same epoch, in the same space-time experiment.
           In this case, as explained above, the speed of O in S' should be $\gamma v$, not $v$.
           Consider, however, an object $\tilde{{\rm O}}$ with the same $x'$ coordinate as O that {\it does} have the
          velocity $v$. The separation $L'$ of O and O' in S' is then equal to that between O' and $\tilde{{\rm O}}$.
          at the epoch of Fig.1b. Compare now the configuration of O and O' in S, with separation $L$
          with the corresponding one  of $\tilde{{\rm O}}$. and O' in S' with separation $L'$. From the symmetry of the
           configurations it can be seen that both $L$ and $L'$ can depend only on $v$:
           $L = L(v)$,  $L' = L'(v)$. The reciprocity of the two configurations is now invoked to
            give the condition, as stated by Pauli~\cite{Pauli}:
            \par{ \tt The contraction of length at rest in S' and observed from S is equal to \newline  the length at
            rest in S 
             as observed from S'}.
            \par       
             The `length at rest in S' ' is $L'$ which `as observed from S' is $L$, whereas
 the `length at rest in S ' is $L$ which `as observed fron S' ' is $L'$. Denoting the contraction factor
                  by $\alpha(v)$, the above condition states that
             \[ L = \alpha(v) L',~~~ L'  = \alpha(v) L \]
         which implies that $L = \alpha(v)^2 L$ or $\alpha(v)^2 = 1$ so that $L = L'$ and the spatial separation
         between O and O' is the same in S and S' at corresponding epochs. The same conclusion is more simply
        reached by noting the symmetry of the configurations of O,O' in S and  $\tilde{{\rm O}}$,O' in S'. and applying
        Leibnitz' Principle of Sufficient Reason~\cite{JHFSTLAWS}.
         \par If, therefore, in the primary experiment, shown in S in Fig.2b and S' in Fig.2a,
         the configuration in S' in Fig.2a is to correctly represent that corresponding to
         the configuration in S in Fig.2b, the velocity $v$ in S' should be replaced by $\gamma v$, so that when 
          PL' flashes O' is aligned with MR(20) in both S and S'. In the reciprocal experiment, shown
         in S in Fig.2a and S' in Fig.2b, $v$ in S
          in Fig.2a should be replaced by $\gamma v$ so that O is aligned with MR'(0) in both S and S'
            when PL flashes. 
           \par Similarly, in the thought experiment of Fig.5, if the S' frame configurations
            on the right side of the figure are to represent observations in this frame of events
             shown in S by the configurations on the left side, instead of what are
              actually shown which are {\it configurations of the physically independent
                reciprocal experiment}, $v$ should be replaced by  $\gamma v$ in all the S' frame
                configurations. In this case, there is no mismatch between the spatial position of
                the decay event in the two frames and the claimed `length expansion' effect does
             not occur. Indeed the claimed `... different PMC corresponding to observations of the light
          flashes emitted by
   PL and PL' in different  frames in (5.7)-(5.10)' is not only `...deeply perplexing for common-sense concepts of space and 
  time.' it is the absurd (self-contradictory) consequence of assuming, at the same time,
    that length intervals are invariant, time dilatation occurs and the conventional interpretation
       of the Reciprocity Principle holds. In conventional special relativity theory
       time dilatation and the Reciprocity Principle are reconciled by invoking the spurious
        `length contraction' effect $dx_{\ORP\OR} = -\gamma x'_{\OR\ORP}$~\cite{NDMIAT}. so that $v' = -v$.
         The correct physical interpretation of the Reciprocity Principle is actually the
       {\it definition} of the configuration
         in S' of the physically-independent experiment that is reciprocal to the primary one
         specified by the standard configuration of the frames S and S'~\cite{JHFPRSTE,JHFSTP3}.
\pagebreak

\end{document}